# Convective Preheating Enhances Front Propagation in DCPD Frontal Polymerization

M. Vijay Kumar, Saujatya Mandal, Siddhant Jain, Saptarshi Basu, Debashish Das[*]

Department of Mechanical Engineering, Indian Institute of Science, Bengaluru, India

## Abstract

Frontal polymerization (FP) enables rapid curing of thermosets via a self-sustaining thermal wave, but its propagation mechanism can shift dramatically depending on processing conditions. In this study, we investigate the effect of trigger direction and monomer viscosity - controlled via hold time - on the front velocity in frontal ring-opening metathesis polymerization (FROMP) of dicyclopentadiene (DCPD). Our experiments reveal that at low viscosities, bottom-triggered FP fronts propagate significantly faster, ~50% faster front speed compared to top-triggered ones, driven by buoyancy-enhanced convection that preheats the unreacted monomer ahead of the front, that can have important implications for manufacturing applications. However, with increasing hold time, the monomer viscosity rises steeply, suppressing convection and causing the front velocity for top and bottom triggering to converge. This behavior reflects a convection-to-conduction (thermal-diffusion) transition in heat transport during FP. Complementary simulations incorporating buoyancy-driven advection reproduce the observed trends and highlight the importance of fluid flow in front dynamics. These results provide new insight into the coupled thermo-fluid-chemical mechanisms in FP offer strategies to tailor front behavior through viscosity and initiation geometry.

## Introduction

Frontal polymerization (FP) is a rapidly evolving technique for polymer synthesis and thermoset curing, characterized by the self-propagating reaction front driven by the exothermic polymerization reaction. Since its initial demonstration [1–3], FP has emerged as a compelling alternative to traditional batch curing methods [4]. Frontal ring-opening metathesis polymerization (FROMP), particularly of dicyclopentadiene (DCPD), has garnered considerable interest due to the rapid curing times, minimal energy inputs required, and exceptional mechanical properties [5, 6]. FROMP of DCPD has found applications across diverse fields, including additive manufacturing [7], composite fabrication [8], self-healing materials [9], and rapid production of lightweight structural materials [5, 10], even for space applications [6].

---

[*] Corresponding Author: ddas@iisc.ac.in

The propagation of frontal polymerization (FP) waves arises from an intrinsic coupling among heat transfer, reaction kinetics, and fluid mechanics. Classical models typically assumed that heat transport within the monomer is purely conductive, thereby neglecting hydrodynamic effects [5, 11]. However, both experimental and numerical investigations have shown that the substantial exothermicity of FROMP systems can generate buoyancy-driven convection within low-viscosity monomers, significantly altering front morphology and velocity [12–14]. These studies demonstrated that natural convection can enhance thermal transport ahead of the front, producing faster, asymmetric, and sometimes non-planar fronts. Despite these insights, prior work has largely focused on convection that emerges during polymerization, while the influence of pre-existing convection (convection resulting from heating DCPD solution before FROMP is triggered) on front initiation and propagation - particularly the stark differences between bottom-triggered (maximally convective) and top-triggered (convection-suppressed) fronts in DCPD - has received little attention. Moreover, the influence of monomer viscosity on the strength of convection - and thus on heat transport and front velocity - remains insufficiently explored, as does the role of the applied heat flux used to trigger the front. Because viscosity governs both fluid flow and thermal boundary-layer structure, it fundamentally determines the relative importance of convective versus conductive transport. Most previous experimental studies have additionally relied on a single hold (incubation or staging) time after catalyst addition, without systematically examining how changes in monomer viscosity during the hold affect convective behavior and front dynamics.

In this study, we directly probe the transition from convection-dominated to conduction-dominated frontal propagation by tuning the pre-polymerization viscosity of DCPD through controlled hold times. Hold time (incubation/staging time), in this work, is defined as the elapsed time between addition of the catalyst to the DCPD mixture and initiation of thermal triggering (start of heating).

By comparing bottom- and top-initiated FROMP fronts across a wide viscosity range, we show that at low viscosities (short hold times), bottom-triggered fronts exhibit markedly higher propagation velocities due to buoyancy-enhanced convection. As viscosity increases (long hold times), convective transport becomes progressively weaker, and top- and bottom-triggered front velocities converge, signifying a shift to a predominantly conductive regime. We further demonstrate that bottom-triggered fronts (heated from below) exhibit a higher propensity for bubble formation and trapped monomer, whereas top-triggered fronts (heated from above) produce more uniform solids - an effect that arises possibly from convection–front coupling and not from differences in chemistry.

To rationalize these observations, we introduce a numerical framework inspired by experimental observation. Our approach first allows convection and conduction to act in the absence of polymerization, mirroring the pre-heating that occurs in bottom-triggered samples. Convection transports heat upward, establishing a vertical temperature gradient that accelerates the subsequent frontal reaction. In contrast, top-triggered systems lack such convective pre-heating. Incorporating this effect into a reaction–diffusion–advection model yields quantitative agreement with measured trends across hold times, highlighting convection as the dominant mechanism distinguishing the two initiation modes.

Overall, findings of this study elucidate the mechanistic interplay among thermal transport, fluid flow, and polymerization kinetics in FP, and demonstrate that controlling monomer viscosity provides a powerful lever for tuning front behavior. These insights offer practical guidelines for optimizing FROMP processes, particularly when uniformity, defect suppression, and controlled front velocities are critical.

## Methods

### Materials

Endo-dicyclopentadiene (DCPD), 5-ethylidene-2-norbornene (ENB), second-generation Grubbs' catalyst (GC2), and phenylcyclohexane (PCH) were supplied by Sigma-Aldrich and used as received. Tributyl phosphite (TBP) inhibitor was sourced from TCI Chemicals.

**Resin Preparation**

The DCPD resin formulation was prepared following the protocol of Robertson et al. [5]. Endo-DCPD, solid at room temperature, was melted at 40 °C, after which 5 wt.% ENB was incorporated to depress the melting point, producing a 95:5 DCPD/ENB blend. This mixture was degassed at 25 kPa for 30 min. Separately, GC2 (monomer:catalyst molar ratio = 10,000:1) was dissolved in PCH and sonicated for 5 min to ensure complete dissolution. TBP was then added in a TBP:GC2 molar ratio of 2:1. The monomer and catalyst phases were combined, stirred, and sonicated for an additional 1 min. All frontal polymerization experiments were performed under standard ambient laboratory conditions.

**Thermal triggering and measurement of front velocity and temperature**

The prepared DCPD formulation was dispensed into flat-bottom 5-mL glass test tubes. Thermal initiation was achieved using an aluminum cylinder containing a central aperture that housed the tip of a soldering iron, enabling controlled heating. The cylinder was positioned to uniformly heat either the top or bottom surface of the test tube. For top-trigger experiments, a glass slide matching the test-tube wall thickness was inserted between the heat source and the sample to replicate the heat-transfer path present during bottom-triggering. Trigger duration was varied between experiments: in one set, heating was discontinued immediately upon visible front initiation, whereas in another, the trigger time was held fixed. For this triggering approach, the heater temperature was ramped at ~0.5 °C s$^{-1}$ to a setpoint of 200 °C. In addition, we conducted a separate set of trials in which the reaction was triggered from the bottom of the test tubes using a butane gas flame.

Front evolution was recorded using a Nikon Z6II digital camera and a Fluke TI480PRO infrared camera. The front velocity was determined from the slope of a least-squares linear fit to front position as a function of time. Front location was extracted either from the refractive-index–induced color change in optical images (analyzed using in-house code and Xcitex ProAnalyst) or from the sharp temperature gradient near the front in the infrared images.

**Measurement of Activation (or Front Initiation) Time**

Activation (front initiation) time was determined using a stopwatch under controlled triggering conditions. The top or bottom heater was maintained at 100 °C using a PID temperature controller. Timing was initiated at the instant the DCPD-filled flat test tube was placed in contact with the heater. To minimize subjective bias associated with visually identifying the onset of front formation, the activation time was operationally defined as the time required for the reaction front to reach a fixed reference position: a marker was placed 17 mm from the triggering end (bottom for bottom triggering; top for top triggering), and the timer was stopped when the front crossed this mark. The recorded elapsed time was reported as the activation time for that experiment.

**PIV Measurements**

Silver hollow spheres were added to the DCPD resin to perform particle image velocimetry (PIV) during frontal polymerization.

**Rheological measurements**

Isothermal rheology was carried out on an Anton Paar MCR 702 rheometer fitted with 25-mm aluminum parallel plates and a solvent trap to minimize evaporation. Time-sweep experiments were performed at 27 °C with a strain amplitude of 0.1% and an oscillation frequency of 1 Hz.

**Heat of reaction and degree of cure analysis**

DSC experiments were conducted on a PerkinElmer DSC 8000 fitted with an intracooling unit. Resin samples were placed in aluminum hermetic pans at room temperature, sealed, and weighed using a Shimadzu ATX 224R analytical balance. Cure behavior of liquid and gelled formulations was characterized between 0 °C and 200 °C at a heating rate of 15 °C/min. Reaction enthalpy was obtained by integrating the baseline-corrected exothermic peak in the heat-flow curve.

More details are included in the Supplemental Material Section.

**Computational modeling**

To account for the impact of convective transport on the polymerization front, the heat transfer and chemical kinetics formulations are integrated with the incompressible Navier–Stokes equations via a Boussinesq treatment[13]. This multiphysics framework allows for the simultaneous evaluation of reaction, diffusion, and convection phenomena within the FROMP system. The set of governing partial differential equations representing this coupled behavior is defined as follows:

**Conservation of Momentum**

$$\rho \left( \frac{\partial \boldsymbol{u}}{\partial t} + \boldsymbol{u}.\nabla \boldsymbol{u} \right) = -\nabla P + \rho \nabla . [\nu(\nabla \boldsymbol{u} + (\nabla \boldsymbol{u})^T)] - \rho \beta \boldsymbol{g} \Delta T \qquad (1)$$

**Mass-conservation**

$$\nabla \cdot \boldsymbol{u} = 0 \quad (2)$$

**Heat Equation**

$$\kappa \nabla^2 T + \rho H_r \left(\frac{\partial \alpha}{\partial t} + \boldsymbol{u} \cdot \nabla \alpha\right) = \rho C_p \left(\frac{\partial T}{\partial t} + \boldsymbol{u} \cdot \nabla T\right) \quad (3)$$

**Cure-Kinetics Equation**

$$\frac{\partial \alpha}{\partial t} + \boldsymbol{u} \cdot \nabla \alpha = A e^{-E/RT}(1-\alpha)^n \quad (4)$$

In this two-dimensional framework, the system is defined by four primary state variables: T(x, z, t), degree of cure, α(x, z, t) and the velocity field $u = [u_x, u_z]$, with (x, z) and t denoting spatial co-ordinates and time. The transition from a liquid monomeric phase to a solid polymer is captured by the scalar field $\alpha \in [0,1]$; this allows for a continuous mathematical representation of the phase change during the advancing front.

To model the fluid dynamics, a buoyancy-induced source term is integrated into the momentum balance using the Boussinesq approximation to account for thermally-induced density gradients. In this context, $\beta$, $g$, and $\Delta T$ signify the thermal expansion coefficient, gravitational acceleration, and the temperature deviation from the reference state, respectively.

In the heat transport equation, $\kappa, C_p, \rho$ & $H_r$ denote the thermal conductivity, specific heat, mass density, and the total exothermic heat of the reaction. The chemical kinetics is governed by an nth order Arrhenius formulation[15] which accounts for autocatalytic behavior via the pre-exponential factor $A$, activation energy $E$, gas constant $R$, and reaction order $n$. Finally, the model assumes that molecular species diffusion is negligible relative to the reaction scales.

The numerical implementation is executed across a 2D rectangular geometry measuring 48 mm in length and 12 mm in width. This computational space is subject to the specific boundary and initial constraints detailed in the subsequent section.

**Initial Conditions on Temperature and Initial Degree of Cure**

$$T(x, z, 0) = T_0 \qquad (x, z) \in \Omega \text{ where } \Omega = \{(x,z) | 0 \leq x \leq W, 0 \leq z \leq H\} \quad (5a)$$

$$\alpha(x, z, 0) = \alpha_0 \qquad (x, z) \in \Omega \quad (5b)$$

**Boundary Conditions (Trigger, Insulated Walls, and No Slip Condition)**

$$-\mathbf{q}\cdot\mathbf{n} = \begin{cases} Q_o & 0 \le t \le t_{stop} \\ 0 & t > t_{stop} \end{cases} \quad (x,z) \in \Gamma_{trig} \quad \Gamma_{trig} = \begin{cases} \{(x,0) | 0 \le x \le W\}, & \text{bottom} \\ \{(x,H) | 0 \le x \le W\}, & \text{top} \end{cases} \quad (6a)$$

$$-\mathbf{q}\cdot\mathbf{n} = 0 \quad (x,z) \in \partial\Omega \setminus \Gamma_{trig} \quad (6b)$$

$$\mathbf{u} = \mathbf{0} \quad (x,z) \in \partial\Omega \quad (6c)$$

Initially, the entire computational domain is prescribed with a uniform ambient temperature $T_0$ and an initial degree of cure $\alpha_0 = 0.2\%$. A heat-flux boundary condition is imposed on a single boundary - either the top or bottom surface, depending on the desired initiation location of frontal ring-opening metathesis polymerization (FROMP) - while the remaining three boundaries are treated as adiabatic. A no-slip condition is enforced along all domain boundaries for the velocity field.

Finite-element simulations of DCPD frontal polymerization were performed using COMSOL Multiphysics. The computational domain was discretized using triangular elements. To accurately capture the sharp gradients in temperature, degree of cure, and velocity across the propagating polymerization front, adaptive mesh refinement was employed based on solution-dependent error indicators.

The simulations were performed as two sequential studies. The first study represents a preheating stage in which cure kinetics are disabled, so the governing equations reduce to coupled heat transfer and buoyancy-driven flow under the imposed thermal boundary conditions. In this stage, convection develops naturally in the bottom-heating configuration due to the unstable thermal stratification, whereas it remains suppressed in the top-heating configuration because the stratification is stable. The second study then captures front initiation and propagation with the reaction kinetics enabled, using the temperature and velocity fields from the preheating stage as initial conditions.

In both studies, time integration was carried out using an implicit backward differentiation formula (BDF) scheme. The nonlinear system was solved using a Newton–Raphson approach: a constant-Newton strategy with Jacobian updates once per time step was used in the second study, while an automatically selected highly nonlinear Newton method was employed in the first study to improve robustness during the early transient regime.

## Results and Discussion

**Trigger Direction Changes Pre-Front Transport**

Figure 1 establishes the central physical contrast between the two initiation modes: top vs bottom heating. As shown in the schematic in Fig. 1(a), in the bottom-triggered configuration, the sample is heated from below and the unreacted monomer layer is susceptible to buoyancy-driven flow; in the top-triggered configuration, heating from above suppresses the density-inversion necessary for sustained convection. The optical time-lapse images show a clear, propagating reaction front in both cases. Some defects can be observed in both triggering cases which will be discussed later.

The accompanying streak images, Fig. 1b, reveal fundamentally different transport fields: Rayleigh–Bénard-type convection [16] is readily observed ahead of the front for bottom-triggered FROMP, whereas no measurable convection is present for top-triggered FROMP. These observations motivate the hypothesis that the measured front velocity is not solely a consequence of reaction kinetics and pure conduction, but is strongly influenced by pre-front heat transport, which is enhanced by buoyancy-driven mixing when triggering from the bottom.

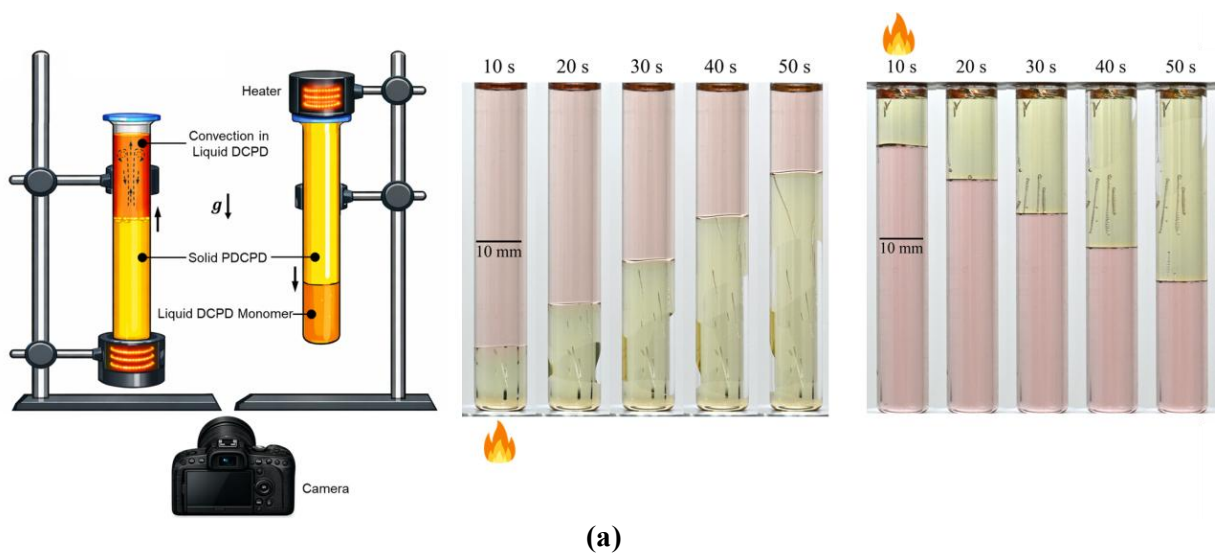

(a)

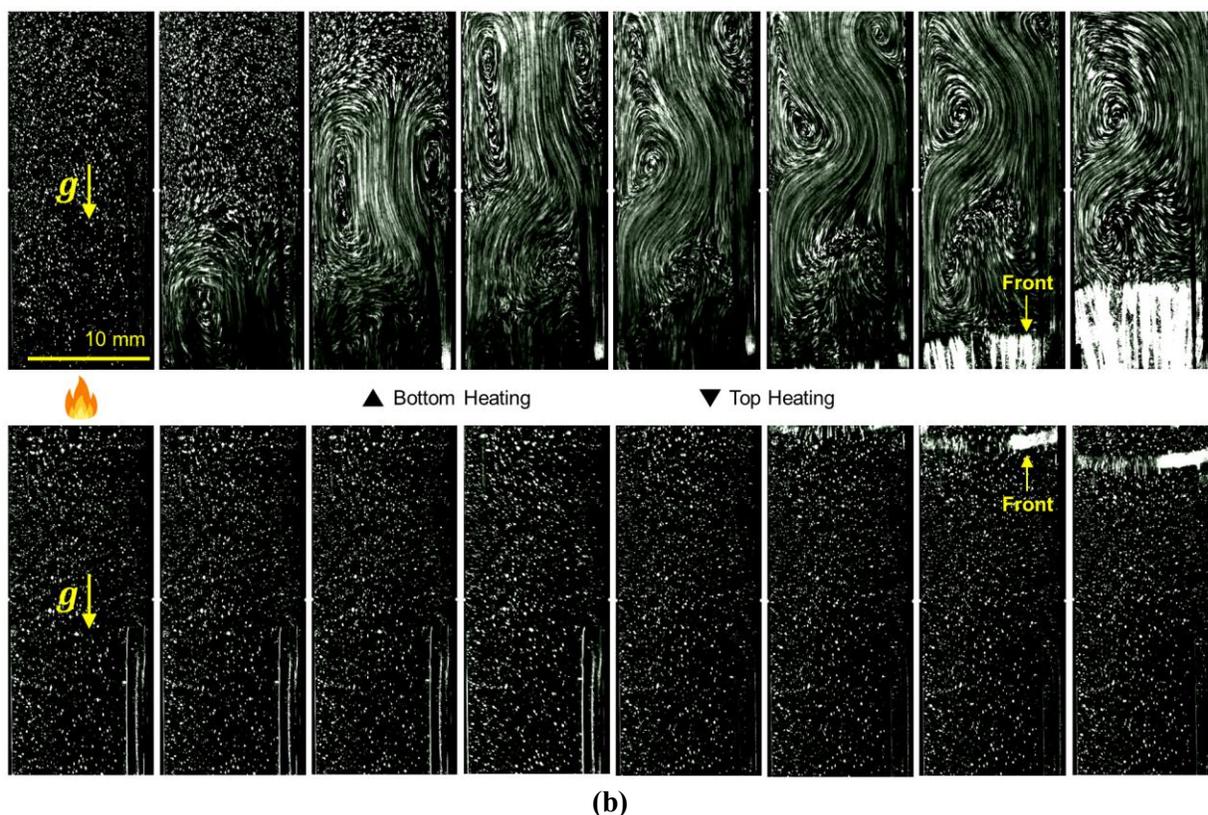

**(b)**

**Figure 1. (a)** (Left) Schematic of the experimental setup used for frontal ring-opening metathesis polymerization (FROMP) of dicyclopentadiene (DCPD). (Middle, Right) Representative time-lapse optical snapshots showing the evolution of the propagating polymerization front for bottom-triggered and top-triggered FROMP, respectively, for a hold time of ~ 900 s in both cases. The time stamps denote the elapsed time from the start of video acquisition, which was initiated manually after the front was first visually detected; they therefore do not represent the absolute time since heating was applied. **(b)** (Top) Streak images constructed from averaged snapshots for a bottom-triggered experiment, illustrating pronounced Rayleigh–Bénard-type convective motion. (Bottom) Corresponding streak images for a top-triggered FROMP experiment, showing the absence of measurable convection. Absolute times are not reported for (b,c) because the field of view was positioned near the upper portion of the tube and recording was manually triggered once the front started near the tube–heater interface.

Upon applying heat at the bottom boundary, the DCPD mixture first undergoes a latency (induction) period during which the bulk remains unreacted while buoyancy-driven convection develops due to an unstable thermal stratification (a warmer, lower-density layer near the heated bottom boundary underlying a cooler, higher-density layer). During this pre-front stage, the temperature field evolves primarily by external heating combined with convective redistribution. The polymerization front is considered to initiate when a localized region adjacent to the heated boundary crosses a thermal runaway (autoacceleration) condition, i.e., when the rate of exothermic heat release from ROMP exceeds the net rate of heat removal (conduction to the boundaries and convective/advection transport into the bulk). This corresponds to reaching an effective ignition temperature (and associated local conversion state) that depends on formulation, heat-loss conditions, and the convective state of the liquid. Once this criterion is met, a self-sustaining reaction zone forms and propagates as a sharp front. Additional support for this hypothesis is provided by the measured initiation (activation) times, which are generally higher for bottom-triggered experiments than for top-triggered experiments (Fig. 2a). This behavior is consistent with the idea that, under bottom heating, buoyancy-driven convection develops during the latency period and can enhance heat removal from the near-boundary trigger region by advecting warm fluid into the bulk, thereby delaying the onset of thermal runaway and front formation. As the hold time increases, the mixture viscosity rises, convective motion is progressively suppressed, and the initiation times for bottom- and top-triggered configurations converge, reflecting a transition toward predominantly conductive preheating in both cases. A qualitatively similar reduction in activation time with increasing holding (or staging) time has been reported in Ref. [17], although the heating protocol and the method used to define/measure initiation time differ from those employed in this study.

**Bottom-triggered fronts are faster at low hold times; velocities converge as viscosity increases**

The effect of trigger direction on kinetics is quantified in Fig. 2(b,c). From the front position–time curves (Fig. 2b), the front velocity is obtained by linear fits to the position-time data, demonstrating higher velocities (slopes) for bottom-triggered fronts under short hold-time (low-viscosity) conditions. Across the broader parameter space (Fig. 2c), the front speed decreases with hold time for both initiation modes, and bottom-triggered fronts propagate substantially faster, ~42% higher on average, than top-triggered fronts at low hold times, while the two curves converge at larger hold times as viscosity increases and convection becomes progressively suppressed.

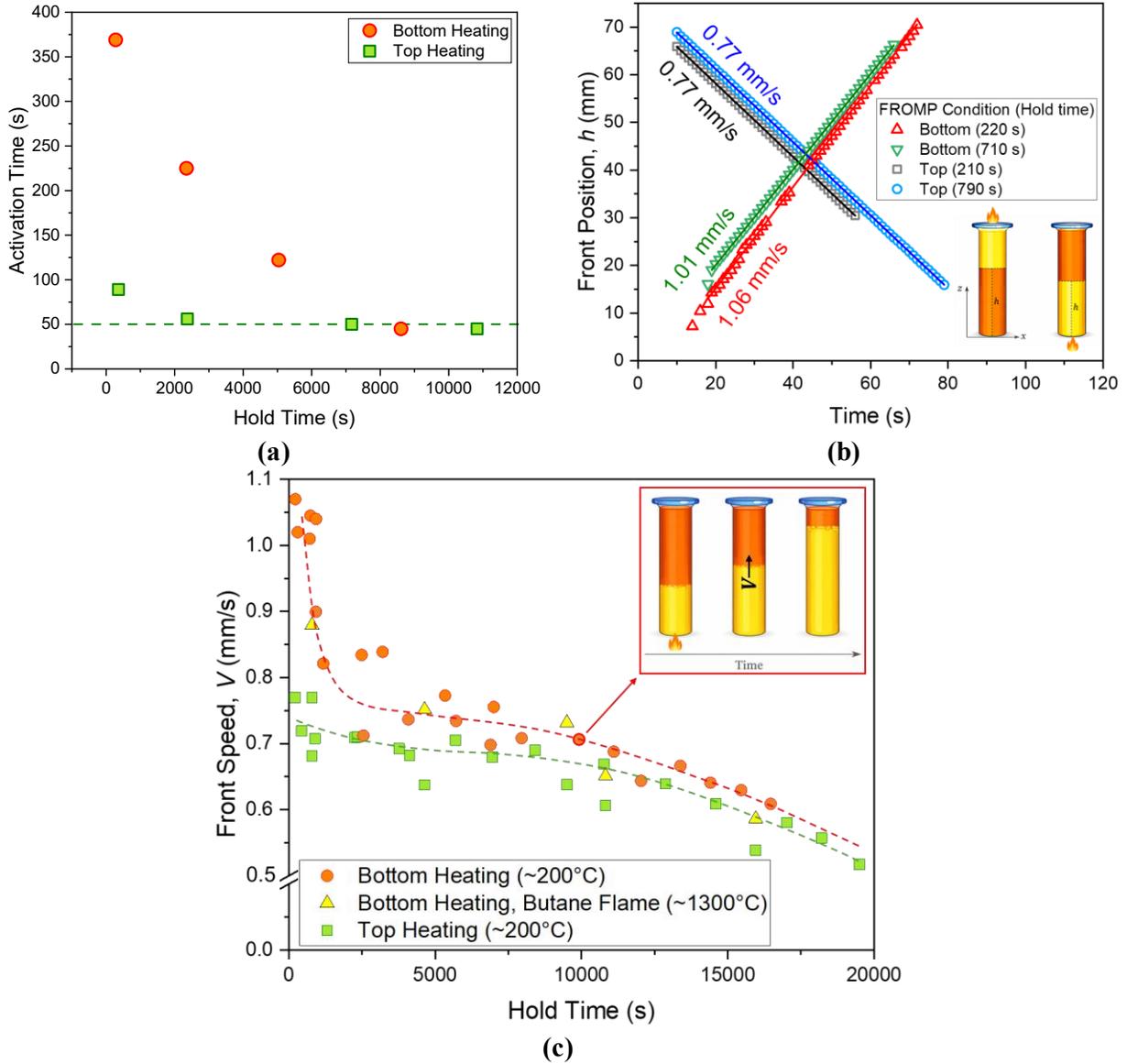

**Figure 2.** **(a)** Activation time as a function of hold time for both bottom and top heating cases; at smaller hold times the activation time for the bottom heating cases are larger than the top heating cases. Also, with increase in viscosity (or hold time), the activation times for top and bottom heating cases converge. **(b)** Front position versus time obtained from post-processed optical images for top-triggered and bottom-triggered FROMP at short hold times. Linear fits to the position–time data yield the front velocity, showing consistently higher speeds for bottom-triggered FROMP under short-hold conditions. The inset schematic shows the coordinate system and the measured front position. **(c)** Front velocity as a function of hold time, highlighting that bottom-triggered fronts propagate substantially faster than top-triggered fronts at low hold times (~ 42%, on average, higher). With increasing hold time (and concomitant increase in mixture viscosity), the velocities converge. Front velocities measured with bottom triggering using either a resistive heater or a butane flame are indistinguishable within experimental scatter, indicating that the higher flame temperature does not measurably increase the propagation speed. It should be noted that each point in the plot is an entire experiment where the test tube is heated from either the bottom (as in the inset schematic) or top and front speed is measured near the center of the test tube. Also, the dashed lines are just for visual purposes.

An additional thing to note is that because hold time is defined as the elapsed time between catalyst addition and the start of thermal triggering, the measured "front velocity at a given hold time" should be interpreted with care. In particular, the front requires a finite time to traverse the test tube length after triggering, and during this propagation interval, the solution can continue to evolve (e.g., gradual conversion and viscosity increase). For the 75 mm travel distance used in this study, the propagation time is approximately 75 s for bottom triggering and ~107 s for top triggering at short hold times. These durations are, however, smaller than the minimum hold time investigated ($\sim$ 200 s), so the reported velocities primarily reflect the material state at the time of triggering and should be regarded as effective (average) velocities over the measurement time window. This averaging is expected to introduce the largest uncertainty at the shortest hold times, where the material state may be changing most rapidly with time. At larger hold times, the propagation time becomes negligible relative to the hold time, and the mixture properties evolve more slowly during the measurement window, making the extracted front velocity more typical of a quasi-steady propagation regime.

An additional and practically important observation is that, for bottom-triggered FROMP, initiating the reaction using a resistive heater (setpoint $\sim$ 473 K) or a butane flame (nominal flame temperature $\sim$ 1573 K) produces indistinguishable front velocities within experimental scatter. This indicates that a hotter external heat source does not necessarily translate into a faster propagating front.

At first glance, this may appear counterintuitive: one might expect that a flame—by imposing a larger thermal driving force—would heat the lower portion of the tube more rapidly, strengthen buoyancy-driven convection, raise upstream temperatures, and thereby increase the front speed. However, the relevant quantity for front dynamics is not the flame's peak temperature, but the effective thermal boundary condition experienced by the reacting mixture (i.e., the heat flux and the interfacial temperature at the tube–mixture boundary). In practice, this effective boundary condition can saturate because heat transfer is limited by (i) thermal resistances through the glass wall and any interfacial contact resistance, and (ii) the rapid formation of a solidified polymer layer near the trigger location. Once a solid layer forms, the trigger region becomes thermally buffered and additional source temperature does not proportionally increase the temperature field within the liquid ahead of the front.

This result is significant because many modeling studies treat initiation through a prescribed trigger temperature $T_{trig}$ as a boundary condition, leading to an apparent sensitivity of predicted front speed to $T_{trig}$. Our experiments show that, under realistic triggering conditions, the front speed is not controlled by the nominal source temperature, but is dominated by reaction–transport coupling within the mixture (conduction and, for bottom triggering at low viscosity, convection) and by the evolving material state (e.g., viscosity and degree of pre-reaction). Thus, a "hotter" trigger does not necessarily yield a faster front once the effective heat transfer into the reactive medium is constrained.

**Infrared thermography directly shows convection-assisted preheating for bottom triggering**

Figure 3 provides thermal-field evidence supporting the mechanism discussed above. For bottom-triggered FROMP at short hold times, thermal images and axial line profiles show elevated temperatures well ahead of the front, particularly toward the upper region of the tube, consistent with convective heat redistribution and preheating of the unreacted monomer. As hold time increases ($0 \to 2 \to 4$ hours), the line profiles show a systematic reduction in this upstream thermal "lift," consistent with increasing viscosity suppressing convection and reducing convective thermal transport.

In contrast, for top-triggered FROMP, the upstream (unreacted) temperature profiles are described as nearly identical ahead of the front across hold times, consistent with the absence of convection and a predominantly conductive thermal field. Notably, the temperature profiles also indicate that peak/front temperatures are higher at lower hold times for both triggering directions, consistent with greater available monomer (lower pre-cure/degree of cure) producing a larger exotherm, whereas partial self-curing at longer hold times reduces available monomer and lowers heat release.

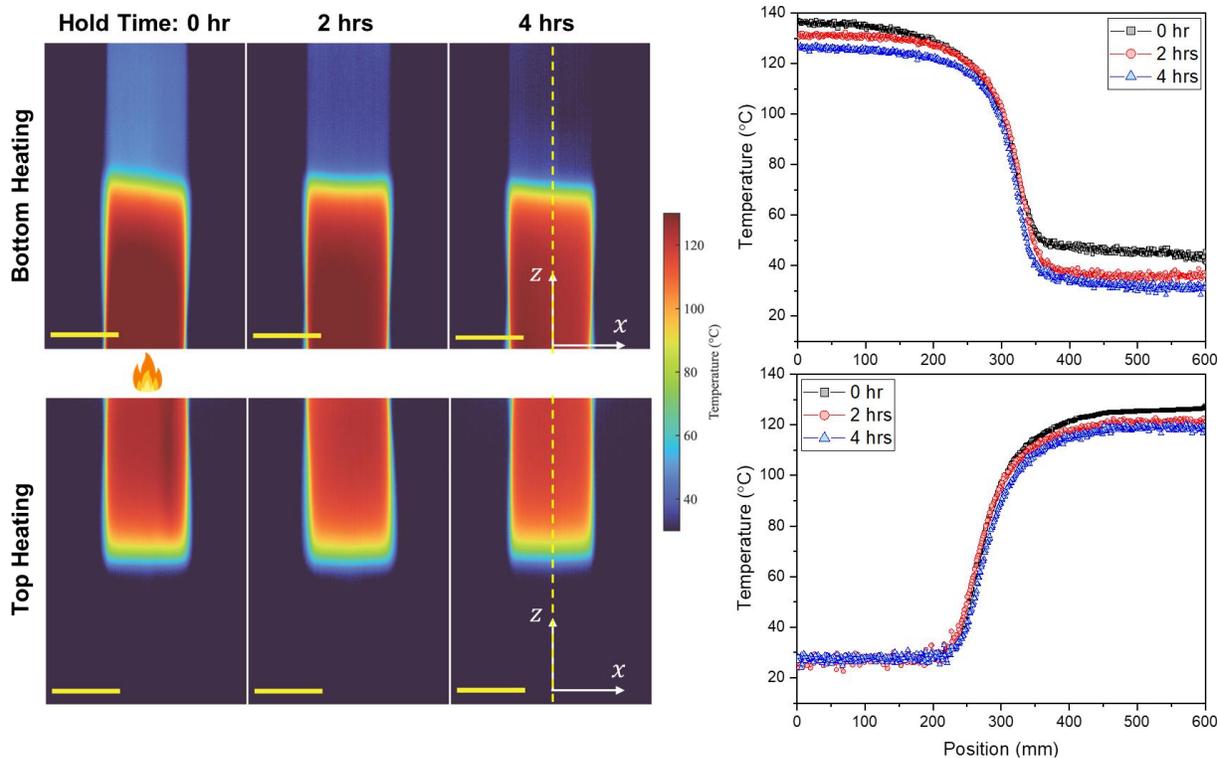

**Figure 3.** Infrared (thermal camera) images of the propagating front for bottom-triggered and top-triggered FROMP at three hold times (0, 2, and 4 h). Scale bar represents 10 mm. The corresponding axial line-temperature profiles extracted along the tube centerline are shown in the plots on the right. For bottom-triggered FROMP at short hold times, elevated temperatures are observed well ahead of the front, particularly in the upper portion of the tube, consistent with Rayleigh–Bénard-type convection enhancing upward heat transport and preheating the unreacted monomer. As hold time increases, the mixture viscosity rises, suppressing convection and progressively reducing thermal transport ahead of the front. In contrast, top-triggered FROMP exhibits no evidence of convective heat redistribution, leading to nearly identical temperature profiles in the unreacted region ahead of the front across hold times. Notably, the peak/front temperatures (behind the front) are higher at lower hold times for both triggering directions, consistent with greater available monomer (lower pre-cure/degree of cure) producing a larger exotherm; with increasing hold time, partial self-curing reduces the available monomer and thus lowers the heat released during FROMP.

It should be noted that the infrared (IR) thermography images report the apparent temperature of the external glass surface of the test tube, not the true temperature field within the reacting DCPD. The measured signal depends on the glass surface emissivity and the radiometric calibration of the camera, and the polymerization front is observed indirectly through heat conducted to the wall. Nevertheless, because all measurements are performed on the same tube material and geometry under identical imaging conditions, the IR data provide a reliable comparative assessment of how the thermal field evolves between top- and bottom-triggered experiments and across hold times. We therefore interpret the systematic differences in the surface temperature patterns as reflecting genuine differences in heat generation and heat transport associated with the FROMP process.

These measurements support the following mechanism: in bottom-triggered FROMP at low viscosity, buoyancy-driven convection redistributes heat from the heated/active region and preheats the monomer above the front, producing a finite axial temperature gradient in the unreacted region. As a result, the front propagates into a warmer medium than in the convection-suppressed (top-triggered) case. Prior studies of frontal polymerization have shown, experimentally [17] and numerically [18], that increasing the initial or ambient temperature increases front velocity by reducing the preheating requirement ahead of the reaction zone and increasing the local reaction rate. In the present DCPD/GC2 system, our results extend this concept by demonstrating that internal convective transport, rather than externally imposed ambient heating, can elevate the upstream thermal state and thereby increase the observed front velocity under bottom-triggered conditions.

**Flow topology evolves with viscosity: from multi-roll/vortex shedding to a weak single-cell pattern**

The streak images in Figure 4 provide a direct visualization of how viscosity (via hold time) modulates the hydrodynamic state. At lower hold time (Fig. 4(a)), bottom-triggered FROMP exhibits two distinct recirculating cells and signatures consistent with vortex shedding, indicating vigorous buoyancy-driven convection in the low-viscosity mixture.

At 2 h hold time (Fig. 4(b)), the bottom-triggered case shows markedly reduced convection with a single localized vortex near the bottom, consistent with viscosity-damped flow and a transition toward diffusion/conduction-dominated heat transfer.

The top-triggered cases do not show comparable convective structures in either hold-time condition, reinforcing that trigger direction controls the stability of buoyancy-driven motion.

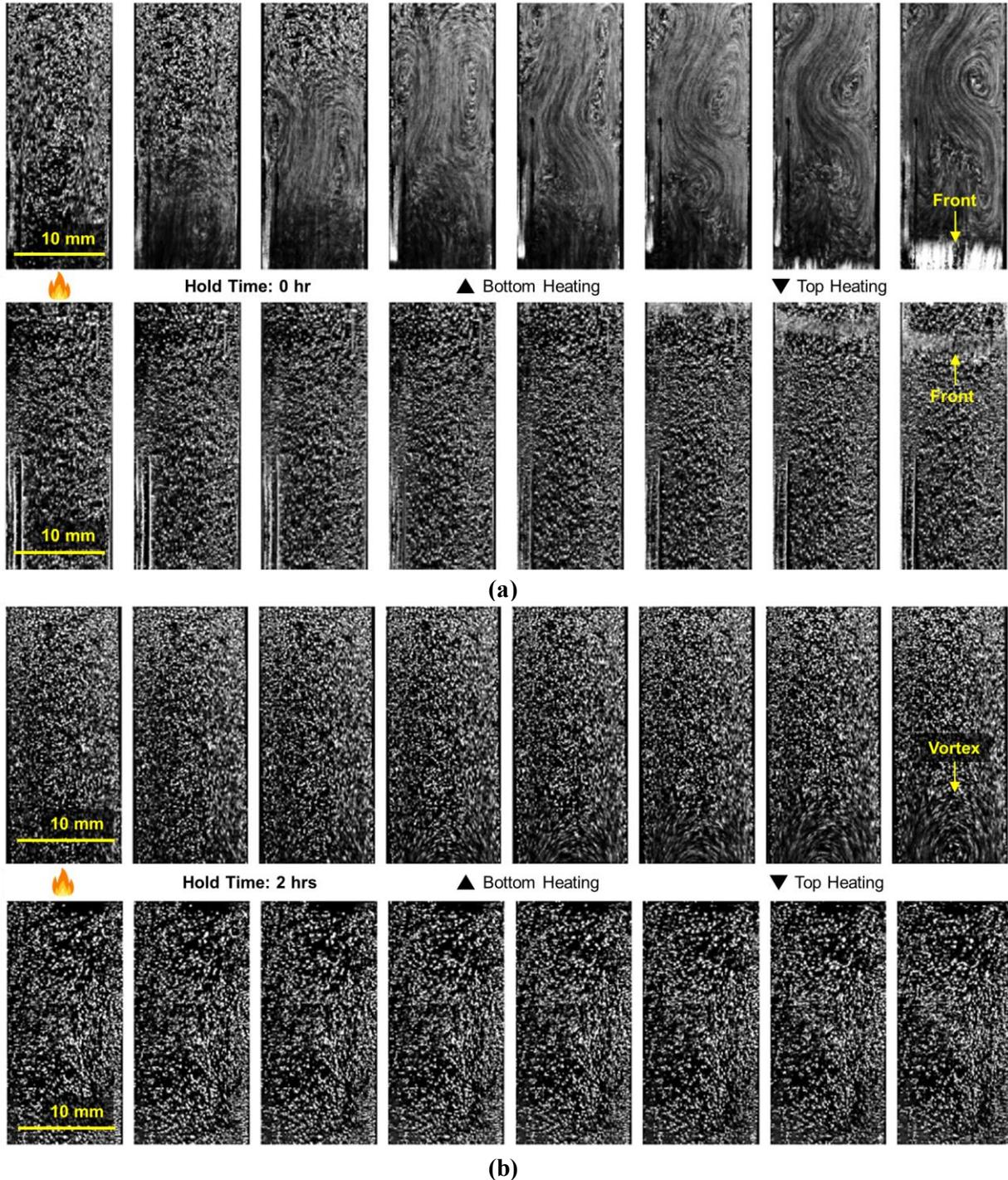

**Figure 4.** Streak images (averaging 200 images in each snapshot) during FROMP for bottom-triggered (top row) and top-triggered (bottom row) experiments at **(a)** 0 h hold time and **(b)** 2 h hold time. Bottom-triggered FROMP exhibits pronounced Rayleigh–Bénard-type convective motion ahead of the advancing front, whereas no such convection is observed for top-triggered FROMP. Increasing hold time (and thus mixture viscosity) suppresses convective activity in the bottom-triggered case. At 0 h hold time, the low-viscosity mixture shows two distinct recirculating cells and evidence of vortex shedding; at 2 h hold time, convection is markedly reduced and a single localized vortex near the bottom is predominantly observed.

**Reaction–transport modeling reproduces the observed convergence trend**

Figure 5 shows that the computational framework reproduces the key experimental trends: simulations that represent top- versus bottom-triggered boundary conditions predict front velocities that agree quite reasonably with measurements across hold time. In the model, hold time is incorporated phenomenologically by varying the initial degree of cure, $\alpha_0$, thereby representing the progressive pre-reaction (and associated changes in mixture state) that occurs during aging prior to ignition. The initial degree of cure as a function of time was determined via differential scanning calorimetry (DSC), while the corresponding viscosity evolution was characterized using rheometry, as described in the Methods section. All simulation parameters and the mapping between $\alpha_0$, viscosity, and experimental hold time are provided in the Supplementary Information.

The modeling premise is that, under bottom triggering, the system undergoes an initial interval of pure thermal transport in which buoyancy-driven convection and conduction redistribute heat within the liquid before a self-sustaining polymerization front is established. This transport produces an upstream thermal field with a finite axial temperature gradient, so that the front subsequently propagates into a preheated medium. Under top triggering, the stratification is stable and buoyancy-driven convection is suppressed, leading to a largely conductive preheating field. Capturing this distinction requires a coupled reaction-diffusion-advection description of the temperature and conversion fields.

In practice, directly integrating the fully coupled governing equations from the onset of heating can fail to reproduce the experimentally observed sequence - namely, the appearance of convection prior to the establishment of a robust propagating front in bottom-triggered experiments. To reflect the experimentally observed chronology and to obtain a stable numerical solution, we therefore use a two-step protocol. Step 1 (preheating): cure kinetics are disabled and the liquid is heated for a certain amount of time $t_{\text{stop}}$ (Fig. 5(a)), allowing the buoyancy-driven flow field to develop under bottom heating (while remaining absent under top heating). Step 2 (reaction): cure kinetics are then enabled, and the front forms and propagates on the pre-established temperature field. Under bottom heating, the preheating step produces a sharper upstream thermal gradient and leads to an initial rapid advancement ("jump") of the reaction zone followed by a stable propagation regime, from which we extract the reported front velocity—consistent with the experimentally observed early-time acceleration.

While the model successfully reproduces the experimentally observed front velocity–hold time trends, several aspects of the coupled physics are necessarily simplified. In the experiments, partial conversion can occur during the latency period, with the mixture viscosity evolving concurrently alongside the development of convective motion. In contrast, in the present two-step implementation in the computational model, the degree of cure is intentionally held fixed during the preheating stage to isolate the effect of transport-induced preheating on subsequent front dynamics. Moreover, the simulations are conducted in two dimensions, whereas the experiments - and the associated buoyancy-driven flow fields - are inherently three-dimensional. Additional simplifying assumptions include adiabatic boundary conditions at the tube walls and the imposition of a finite-duration heat flux that is set to zero after $t_{stop}$, both of which deviate from experimental conditions. These approximations can influence the predicted absolute front velocities and the detailed flow topology, even though the model captures the correct parametric trends.

Despite these limitations, the model reproduces the salient thermal signatures observed experimentally. The normalized temperature-distance profiles (inset) show qualitative agreement with measurements and, importantly, recover the same trends: (i) in bottom-triggered cases, convection effectively redistributes heat and yields elevated temperatures ahead of the front, whereas no comparable upstream preheating is present for top triggering; and (ii) for both trigger directions, the temperature at (and behind) the front decreases with increasing hold time (equivalently, higher $\alpha_0$), consistent with reduced available monomer and diminished exothermic heat release. The remaining differences in the exact temperature magnitudes are attributed to the simplified viscosity evolution, preheating protocol and the reduced dimensionality of the simulations.

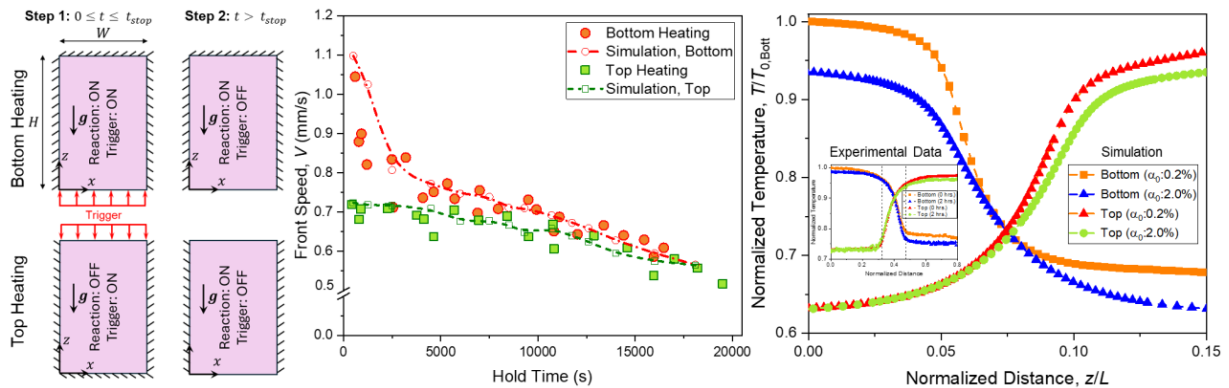

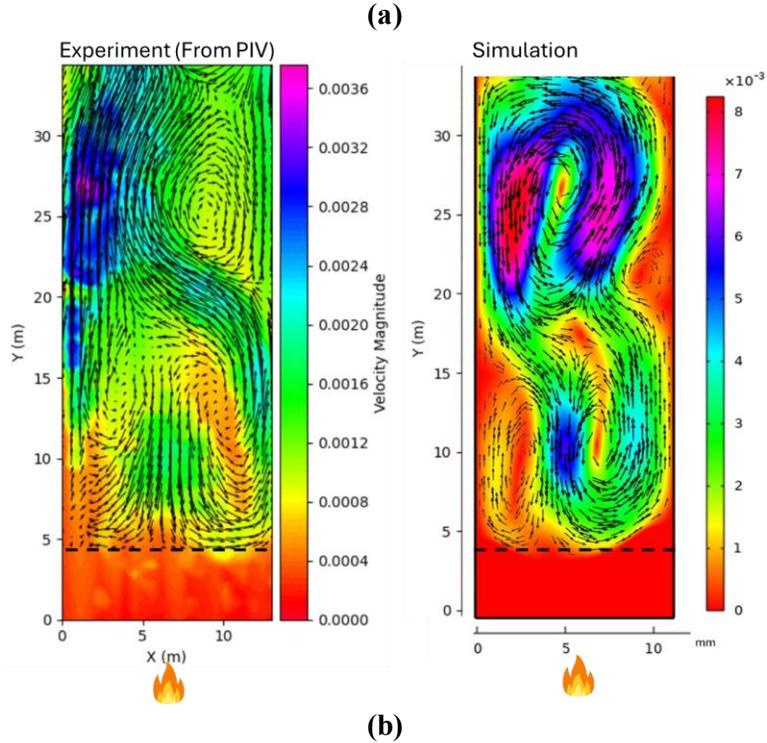

**Figure 5. (a)** Left: Schematic of the computational setup and the boundary conditions. Middle: Experimental and computational front velocities vs hold time match reasonably. Right: Normalized temperature as a function of normalized distance showing reasonable match with experimental results (inset), and more importantly providing the same trends as the experiments (temperature is effectively transported by convection in the bottom heating case leading to higher temperature ahead of the front (not the case for top heating); temperature at (or behind) the front decreases with hold time (or higher $\alpha_0$) for both top and bottom heating **(b)** The velocity magnitudes of the fluid due to convection matches reasonably for the experiments and simulation, though the difference can be attributed to the fact that the simulations are only performed on 2D rather than the 3D nature of the experiments.

**Defect Formation and Convection-Front Coupling**

Optical imaging shows that defect formation can occur under both triggering configurations, including top-triggered FROMP (Fig. 1). However, the incidence, size scale, and morphology of the defects differ systematically between top and bottom triggering (Fig. 6). Top-triggered specimens are generally more uniform and, when defects are present, they typically appear as isolated bubbles or small localized heterogeneities. In contrast, bottom-triggered specimens exhibit a markedly higher propensity for extended, flow-aligned defect structures, including long curvilinear streaks that, upon closer inspection, resolve into discrete bead-like trains following a curved trajectory.

This qualitative distinction is consistent with the transport regimes established during the latency period. Under bottom triggering, heating from below produces an unstable density stratification and drives Rayleigh–Bénard-type convection prior to, and during the earliest stages of, front formation (Figs. 1 and 4). Even when the formulation is degassed, small inclusions can still arise from residual nuclei introduced during handling, heterogeneous nucleation at the wall, or the growth of trace microbubbles promoted by local heating and/or volatile impurities. In the presence of convection, such inclusions are not trapped randomly; instead, they are advected and repeatedly swept along preferred streamlines associated with convective plumes and recirculation cells. As the polymerization front advances and viscosity rises sharply behind the front, these moving inclusions can be captured and immobilized before they coalesce or vent, leaving the observed void-streak / bead-train morphology. The quasi-periodic "beads-on-a-string" appearance, Fig. 6, further suggests that the underlying flow is at least partially time-dependent (e.g., plume oscillations or vortex shedding) and/or that inclusions undergo repeated breakup/pinch-off under shear and extensional components of the convective field before being frozen in place.

Under top triggering, buoyancy-driven convection is strongly suppressed by the stably stratified thermal field (Figs. 1 and 4). Consequently, inclusions that do form are less likely to be organized into extended trains and are more likely to remain isolated, coalesce locally, and/or migrate upward toward the free surface without being swept into coherent plume trajectories. This provides a natural explanation for why defects are still possible in top-triggered runs (Fig. 1), yet the prominent streak-like defects are far less frequent than in bottom-triggered runs (Fig. 6). In this sense, the observed differences arise primarily from convection–front coupling and the associated inclusion-transport pathways enabled (or suppressed) by the triggering geometry.

Notably, we also find that these streak/bead-train defects are not always observed in otherwise identical bottom-triggered experiments conducted without TBP, despite the presence of convection. This indicates that convection alone is unlikely to be sufficient to generate the observed defect morphology; rather, convection appears to amplify a defect-seeding mechanism that is enabled under TBP-containing formulations. While the origin of this coupling is not yet fully resolved, a plausible hypothesis is that TBP- by attenuating catalyst activity and extending the latency window - renders the local reaction rate and gelation dynamics more sensitive to small temperature or compositional variations that are transported and reinforced by convective plumes. Such TBP-mediated kinetic heterogeneity could facilitate intermittent inclusion formation and subsequent flow-aligned trapping, producing the bead-train morphology observed under bottom triggering. We emphasize that this proposed link to TBP is qualitative at present and motivates future systematic tests (e.g., TBP-concentration sweeps coupled with defect quantification).

Finally, the defect density and severity decrease with increasing hold time, particularly for bottom triggering (Fig. 6). This trend is consistent with viscosity growth suppressing convective motion: as viscosity increases, plume strength and recirculation weaken, reducing both the ability of the flow to entrain/align inclusions and the likelihood that inclusions are transported into the gelation zone in a manner that produces extended streak defects. Thus, as the system transitions toward a more conduction-dominated preheating regime, defect formation becomes less strongly amplified by internal flow.

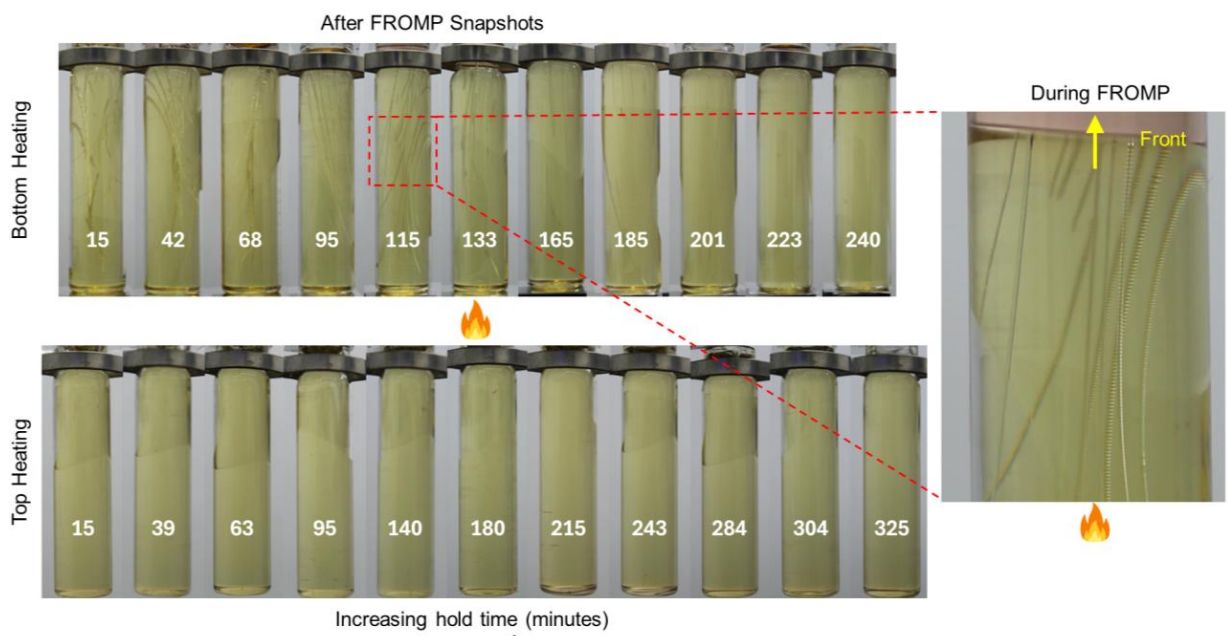

**Figure 6.** Trigger-dependent defect formation in solid DCPD. Representative cured specimens obtained after bottom-triggered (top row) and top-triggered (bottom row) frontal polymerization under otherwise identical formulation and processing conditions. Bottom-triggered samples exhibit prominent flow-aligned streak defects that, upon magnification (right inset), resolve into bead-like inclusion trains following curved trajectories; the yellow arrow marks the approximate location/direction of the propagating front. In contrast, top-triggered specimens are comparatively uniform, with defects-when present-appearing primarily as isolated bubbles or weak localized heterogeneities (Fig.1). The contrast in some top-triggered samples results from polymerization-induced volumetric shrinkage due to increased density upon curing.

## Conclusion

In this study we show that the propagation of frontal ring-opening metathesis polymerization (FROMP) in DCPD is governed not only by reaction kinetics and thermal diffusion, but also - critically - by the transport regime established during the latency period, which is strongly controlled by trigger direction and the time-dependent viscosity of the reacting mixture. Bottom triggering produces an unstable thermal stratification that drives buoyancy-induced convection prior to front formation, whereas top triggering suppresses convection and yields predominantly conductive preheating. By combining time-resolved optical tracking of front position, infrared thermography, and PIV-based flow visualization, we demonstrate that bottom-triggered fronts propagate substantially faster than top-triggered fronts at short hold times, when the mixture is least viscous and convective transport is strongest. With increasing hold time, viscosity increases, convective motion is progressively damped, and the front velocities for top and bottom triggering converge, consistent with a transition to conduction-dominated preheating in both configurations. Infrared measurements further reveal that bottom triggering leads to elevated temperatures ahead of the front, consistent with convective redistribution of heat, while top triggering exhibits little to no upstream preheating. A key control experiment shows that, under bottom triggering, initiation with either a resistive heater or a butane flame yields indistinguishable front velocities within experimental scatter. This implies that the nominal source temperature does not directly determine front speed; instead, propagation is governed by the effective heat transfer into the mixture and subsequent reaction–transport coupling, with implications for models that treat $T_{trig}$ as the sole boundary control on velocity. Finally, a reaction-diffusion-advection computational framework, implemented with a two-step preheating–reaction protocol consistent with the experimentally observed sequence (convection preceding front formation), captures the principal experimental trends in front velocity and normalized temperature profiles across hold times. While the simulations simplify certain aspects-most notably the coupled evolution of conversion/viscosity during the latency period, the boundary conditions, and the inherently three-dimensional nature of buoyant flow—they provide mechanistic support for the central conclusion: convective preheating under bottom triggering accelerates front propagation when the mixture is sufficiently mobile, and this enhancement diminishes as viscosity increases.

**Data Availability**

Data and code will be provided upon reasonable request.

**Competing Interests**

The authors declare no competing interests.

**Acknowledgements**

The authors gratefully acknowledge the support of the ISRO–IISc Space Technology Cell under Project Code ISTC/MME/DD/485. The authors also extend their sincere thanks to Dr. P. Subba Rao, Group Director of the Structures Group at the U. R. Rao Satellite Centre, ISRO, India, for his valuable discussions and insights.